\documentclass[a4paper,11pt]{article}
\usepackage{pos}
\usepackage{bm}
\usepackage{graphicx}

\title{Femtoscopy study of $\pi^-\Lambda$ and $K^-p$ interactions}

\author[a]{Pablo Encarnación}
\author[b]{Albert Feijoo}
\author[a]{Àngels Ramos}

\affiliation[a]{Departament de F\'{i}sica Qu\`antica i Astrof\'{i}sica and Institut de Ci\`encies del Cosmos (ICCUB), Facultat de F\'{i}sica, Universitat de Barcelona, Barcelona, Spain}

\affiliation[b]{Physik Department E62, Technische Universit\"at M\"unchen, Garching, Germany, EU}

\abstract{We have calculated the femtoscopic correlation functions of meson-baryon pairs in the strangeness $S=-1$ sector, employing a unitarized chiral interaction model up to next-to-leading order. We will show preliminary results for the $\pi^-\Lambda$ correlation function, which is presently under analysis by the ALICE@LHC collaboration. We will also demonstrate that, within $2\sigma$, the employed interaction is perfectly capable of reproducing the $K^-p$ correlation function data measured by the same collaboration, without the need of changing the coupled-channel interactions, as has been suggested recently.}

\FullConference{10th International Conference on Quarks and Nuclear Physics (QNP2024)\\
8-12 July, 2024\\
Barcelona, Spain\\}

\begin{document}
\maketitle

\section{Introduction}

At low energy, effective field theories, and in particular Unitarized Chiral Perturbation Theory (UChPT), replace QCD's standard perturbative methods in describing the interactions among hadrons. Traditionally, scattering experiments have been the main source to constrain such theories, but in the last decade hadron femtoscopy has surfaced as a valuable alternative. In this work we study the femtoscopic correlation function (CF) of $S=-1$ meson-baryon (MB) pairs in coupled-channels (cc), employing UChPT up to NLO. We revisit the $K^-p$ interaction explored in \cite{Alice2020} and provide a novel prediction for the $\pi^-\Lambda$ CF, currently under analysis by the ALICE collaboration.

\section{Formalism}

In the $S=-1$ sector, different MB channels with the same quantum numbers couple to each other. In particular, for the $Q=0$ block we consider 10 cc ($K^-p$, $\bar{K}^0n$, $\pi^0\Lambda$, $\pi^0\Sigma^0$, $\eta\Lambda$, $\eta\Sigma^0$, $\pi^+\Sigma^-$, $\pi^-\Sigma^+$, $K^+\Xi^-$, $K^0\Xi^0$), and for $Q=-1$ we consider 6 cc ($ \pi^-\Lambda$, $\pi^0\Sigma^-$, $\pi^-\Sigma^0$, $K^-n$, $\eta\Sigma^-$, $K^0\Xi^- $). The effective chiral Lagrangian up to NLO, $\mathcal{L_{\phi B}} = \mathcal{L}^{(1)} + \mathcal{L}^{(2)}$, is given by
\begin{eqnarray}
    \mathcal{L}^{(1)} &=& \langle \bar{B} i \gamma^\mu D_\mu B \rangle - M_0 \langle \bar{B}B \rangle + \frac{1}{2} D \langle \bar{B}\gamma^\mu\gamma^5 \{ u_\mu,B \} \rangle + \frac{1}{2} F \langle \bar{B}\gamma^\mu\gamma^5 [ u_\mu,B ] \rangle \, ,
    \label{eq:lagrangian1}
\end{eqnarray} \vspace{-1.4cm}

\begin{eqnarray}
    \mathcal{L}^{(2)} &=&  b_D \langle \bar{B} \{\chi_+,B\} \rangle + b_F \langle \bar{B}[\chi_+,B] \rangle + b_0 \langle\bar{B}B\rangle \langle \chi_+ \rangle + d_1 \langle\bar{B}\{u_\mu,[u^\mu,B]\} \rangle  \nonumber \\
    &+& d_2 \langle \bar{B}[u_\mu,[u^\mu,B]] \rangle + d_3 \langle \bar{B}u_\mu \rangle \langle u^\mu B \rangle  + d_4 \langle \bar{B}B \rangle \langle u^\mu u_\mu \rangle \, .
    \label{eq:lagrangian2}
\end{eqnarray}
Here, $B$ is the octet baryon matrix, and the matrix of pseudoscalar mesons $\phi$ is contained in the field $u_\mu = i u^\dagger \partial_\mu U u^\dagger$, with $U = u^2 = \exp(i \sqrt{2} \phi / f)$. The covariant derivative is defined as $D_\mu B = \partial_\mu B + [\Gamma_\mu, B]$, where $\Gamma_\mu = (u^\dagger \partial_\mu u + u \partial_\mu u^\dagger)/2$, and $\chi_+=-\{\phi,\{\phi,\chi\}\}/4f^2$, with $\chi=\textrm{diag}(m_\pi^2,\ m_\pi^2,\ m_K^2-m_\pi^2)$. This Lagrangian depends on a set of parameters referred to as the Low Energy Constants (LECs) ($f,D,F,b_D,b_F,b_0,d_1,d_2,d_3,d_4$).

The s-wave component of the interaction kernel, $V_{ij}$, derived from the chiral Lagrangian, must be unitarized in order to dynamically generate resonances. This is done by solving the Bethe-Salpeter (BS) equation. In the on-shell approximation \cite{Oset:onshell,Hyodo:onshell}, the BS equation, and thus the scattering amplitudes, can be expressed in the matricial form  $T = (1-VG)^{-1}V$, where $G$ is a diagonal matrix whose elements are the MB loop functions, which after using dimensional regularization are \cite{Oller:dimensional} 
\begin{eqnarray}
    G_l(\sqrt{s}) = \frac{2M_l}{16\pi^2} \bigg[ a_l(\mu) + \ln\frac{M_l^2}{\mu^2} + \frac{m_l^2-M_l^2 + s}{2s}\ln\frac{m_l^2}{M_l^2} + \frac{q_l}{\sqrt{s}} \ln\frac{(s+2q_l\sqrt{s})^2 - (M_l^2-m_l^2)^2}{(-s+2q_l\sqrt{s})^2 - (M_l^2-m_l^2)^2} \bigg]
\end{eqnarray}
where $M_l$ and $m_l$ are, respectively, the baryon and meson masses of the l-th channel and $q_l$ the center-of-mass momentum of the $l$-th channel pair at a $\sqrt{s}$ energy. The $a_l$ are the so-called substraction constants (SCs), which cure the divergence for a given regularization scale $\mu$. These SCs, together with the LECs, are the free parameters of the formalism and their values are taken from the BCN model \cite{Feijoo:BCN} in the present study.

For the $K^-p$ channel, the Coulomb interaction should be also taken into consideration in addition to the strong contribution, which is implemented in the amplitude as
\begin{eqnarray}
    T_{ij}(\sqrt{s},p,p') = T_{ij}^S(\sqrt{s}) + \delta_{ij} T^c(\sqrt{s},p,p')
\end{eqnarray}
where $T_{ij}^S$ is the strong scattering amplitude and $T^c$ is the Coulomb one, obtained from the s-wave Coulomb interaction kernel applying relativistic correction factors as described in \cite{Torres:coulomb}.

In the forementioned multichannel scenario, the two-particle CF of the observed $i$-th channel can be expressed through the generalized Koonin-Pratt formula \cite{Fabbietti}, 
\begin{equation}
\label{eq:CF}
    C_i(p) = \sum_{j} w_j \int d^3 r \ S_j(r)|\Psi_{ji}(p,r)|^2\,,
\end{equation}
where $p$ and $r$ are the relative momentum and distance between the two particles, respectively. The preceding summation covers all possible inelastic transitions to the final state, and the production weights $w_j$ account for the channel production relative to that of the final state. 

The relative wave function of the particle pair in Eq. (\ref{eq:CF}), $\Psi_{ji}(p,r)$, can be obtained following the detailed explanation in \cite{Vidaña:fdo}. Finally, the source function $S_j(r)$, which represents the probability distribution of producing the $j$-th pair at a relative distance $r$, is parametrized by a spherical Gaussian, $S_j(r)=(4\pi R_j)^2\exp(-r^2/4R_j^2)$, whose size can be channel dependent and subject to the feedback from strongly decaying particles. For the $K^-p$ CF, the employed source sizes are the experimental ones obtained by \cite{Alice2020}: $R_{\bar{K}N}=1.08\pm0.22$ fm and $R_{\pi\Sigma,\pi\Lambda}=1.23\pm0.26$ fm. For the other channels we fix $R_{\eta\Lambda,\eta\Sigma^0,K^+\Xi^-,K^0\Xi^0}=1.2$ fm. For the $\pi^-\Lambda$ CF no experimental information is yet available, and the same source size, $R=1.2$ fm, is used for all the involved coupled channels.

\section{Results}

\begin{figure}[hbt!]
    \centering
    \includegraphics[width=0.8\textwidth]{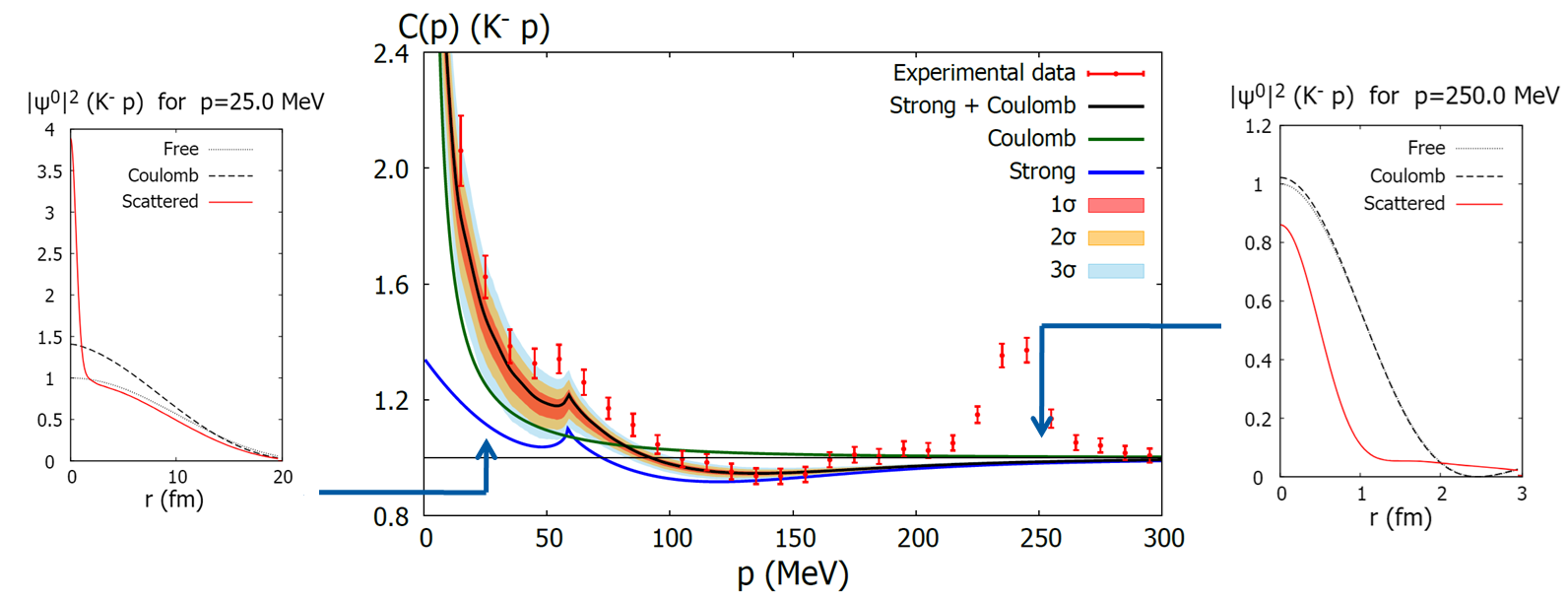}
    \caption{Main plot: $K^-p$ CF and its errorbands obtained from the BCN model parameters. The strong, Coulomb and strong+Coulomb contributions are shown together with the experimental data from ALICE \cite{Alice2020}. Left and right plots: free and scattered spatial wave functions at relative momenta 25 MeV and 250 MeV. }
    \label{fig:k-p}
\end{figure}

Fig. \ref{fig:k-p} shows the $K^-p$ CF predicted by the BCN model together with the experimental data from \cite{Alice2020}. The errorbands of the full CF (black line) are derived from the uncertainties of the parameters of the BCN model. Excluding the $\Lambda(1520)(3/2^-)$, which cannot be reproduced with an s-wave approach, it can be appreciated that the model does a good job reproducing the experimental data, the major discrepancies appearing around the $\bar{K}^0n$ threshold, where the theory understimates the data points. At low momenta, the attractive behaviour of the $\bar{K}N$ interaction that gives rise to the formation of the $\Lambda(1405)$ can be appreciated from the scattered WF of the left plot, while the right plot displays the wavefunction for a large momentum showing the repulsive character of the interaction in this range. \vspace{-0.2cm}

\begin{figure} [hbt!]
    \centering
    \begin{minipage}{0.49\textwidth}
        \centering
        \includegraphics[width=0.75\textwidth]{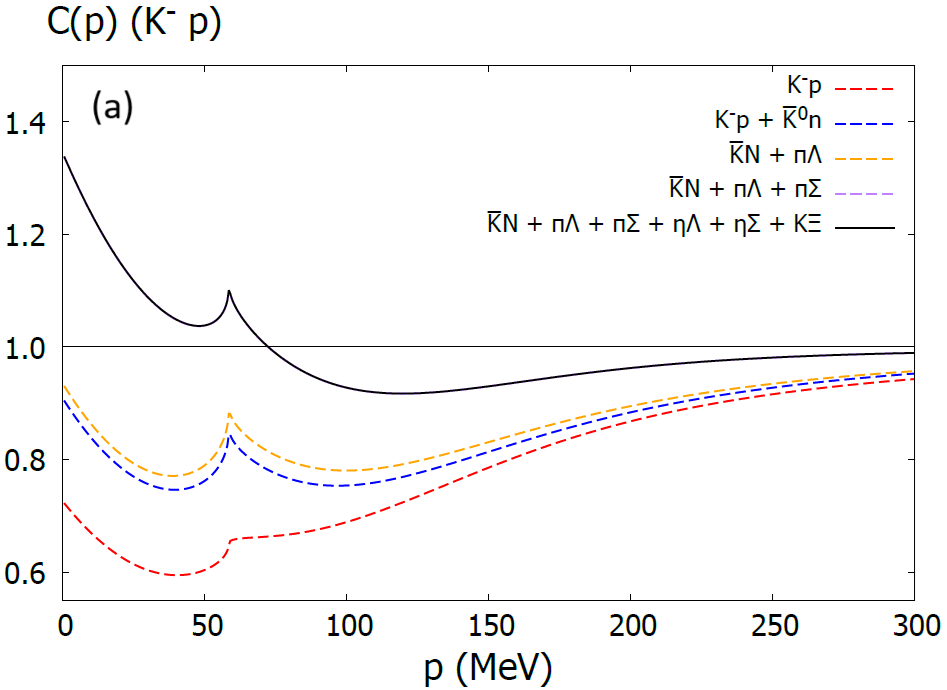}
    \end{minipage}
    \begin{minipage}{0.49\textwidth}
        \centering
        \includegraphics[width=0.75\textwidth]{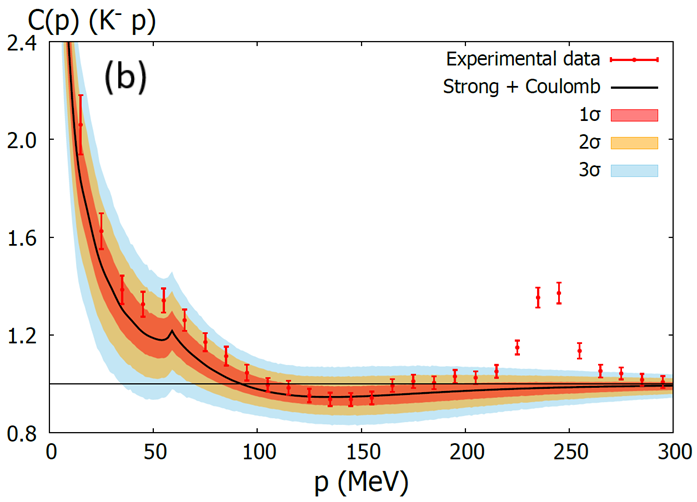}
    \end{minipage}
    \caption{a) Sequential contributions of the coupled channels to the $K^-p$ CF. b) $K^-p$ strong+Coulomb CF and its errorbands obtained from the source radii, together with the experimental data from ALICE \cite{Alice2020}.}
    \label{fig:k-p2}
\end{figure}

Fig. \ref{fig:k-p2}a shows the sequential contributions of the coupled channels to the $K^-p$ strong-only CF are shown. The elastic channel (red line) by itself is unable to produce the full CF, and the contribution of the other channels is needed, specially the $\bar{K}^0n$ and $\pi\Sigma$ ones. Fig. \ref{fig:k-p2}b shows the full CF predicted by the BCN model including also the Coulomb contribution, where the errorbands are now derived from the source radii uncertaintiess. In \cite{Alice2020} it was claimed that a modification of the strength of the coupled-channel interaction was needed in order to correctly reproduce the experimental CF. As shown in this figure, we do not support this claim since the unmodified theory is perfectly able to reproduce the experimental data almost within 1$\sigma$ when the source uncertainties are considered.

\begin{figure}[hbt!]
    \centering
    \includegraphics[width=0.8\textwidth]{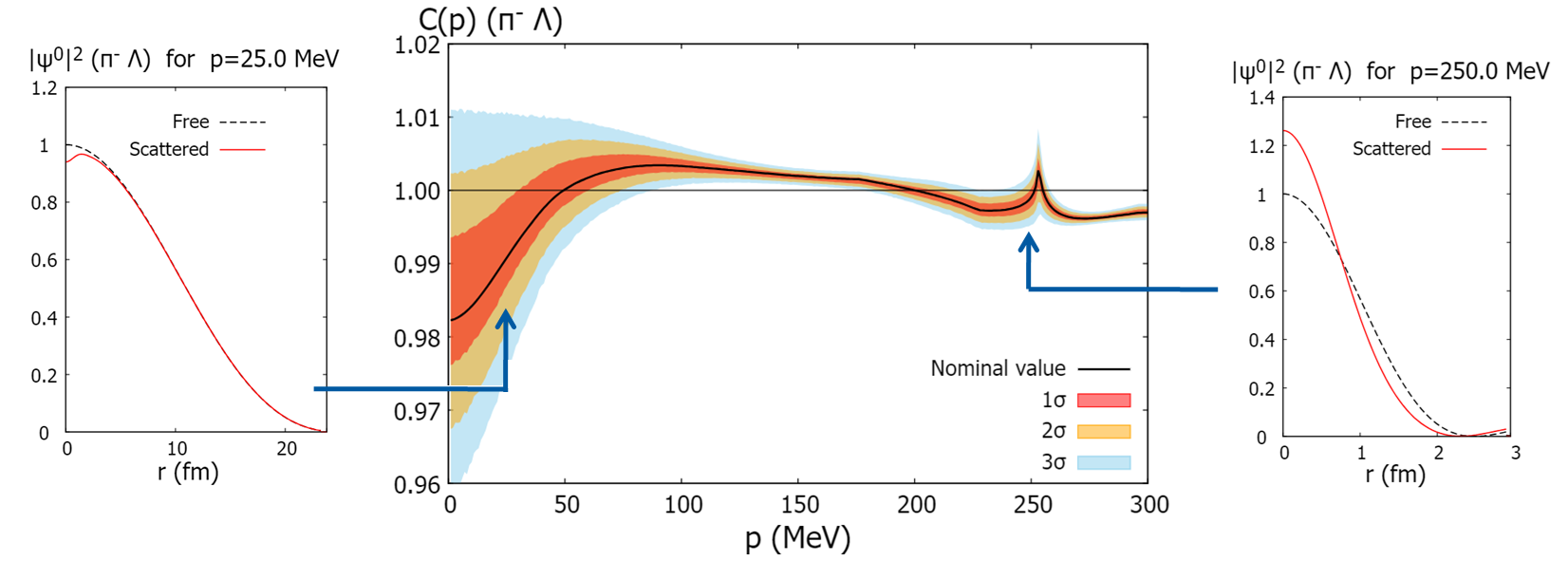}
    \caption{Main plot: $\pi^-\Lambda$ CF and its errorbands obtained from the BCN model parameters. Left and right plots: free and scattered spatial wave functions at relative momenta 25 MeV and 250 MeV, respectively.}
    \label{fig:pi-lambda}
\end{figure}

We now analyze our prediction of the $\pi^-\Lambda$ CF. Even though no experimental information on this channel is yet available, it is currently under analysis by ALICE, and this prediction will serve to test the theory below the $K^-p$ threshold. In Fig. \ref{fig:pi-lambda} the $\pi^-\Lambda$ CF is shown with the errorbands derived from the BCN model. Unlike the $\bar{K}N$ one, the $\pi\Lambda$ strong interaction is very weak, as reflected in the CF strength, and the $K^-n$ threshold can be clearly seen as a cusp at 250 MeV. At low momenta, the scattered WF indicates a very weak repulsive interaction, while at higher momenta the WF is enhanced by an attractive behaviour.

\begin{figure}[hbt!]
    \centering
    \begin{minipage}{0.49\textwidth}
        \centering
        \includegraphics[width=0.75\textwidth]{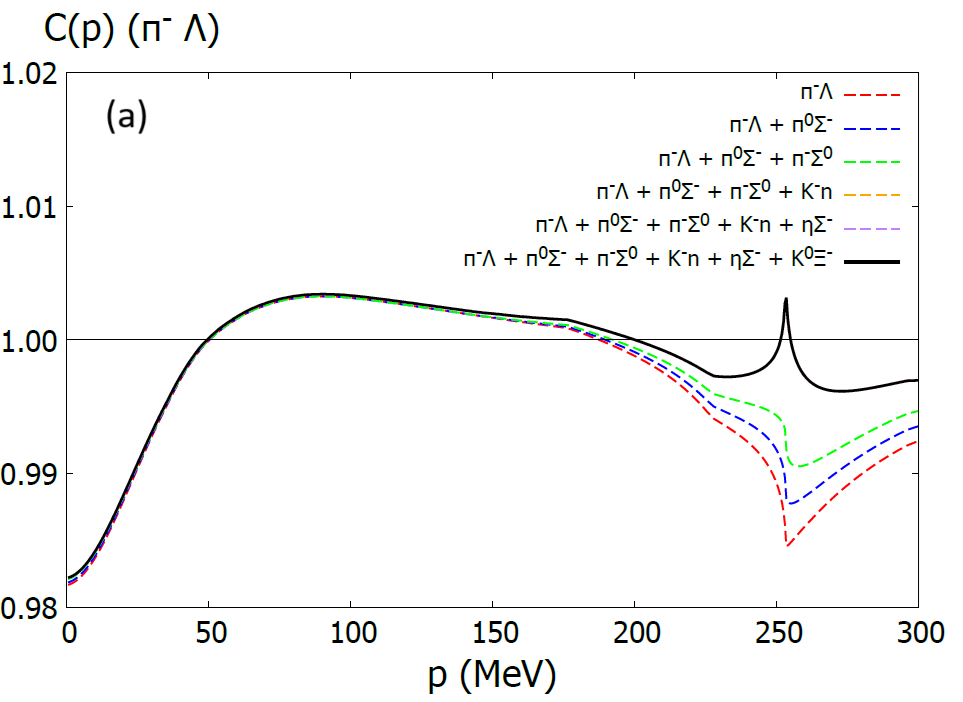}
    \end{minipage}
    \begin{minipage}{0.49\textwidth}
        \centering
        \includegraphics[width=0.75\textwidth]{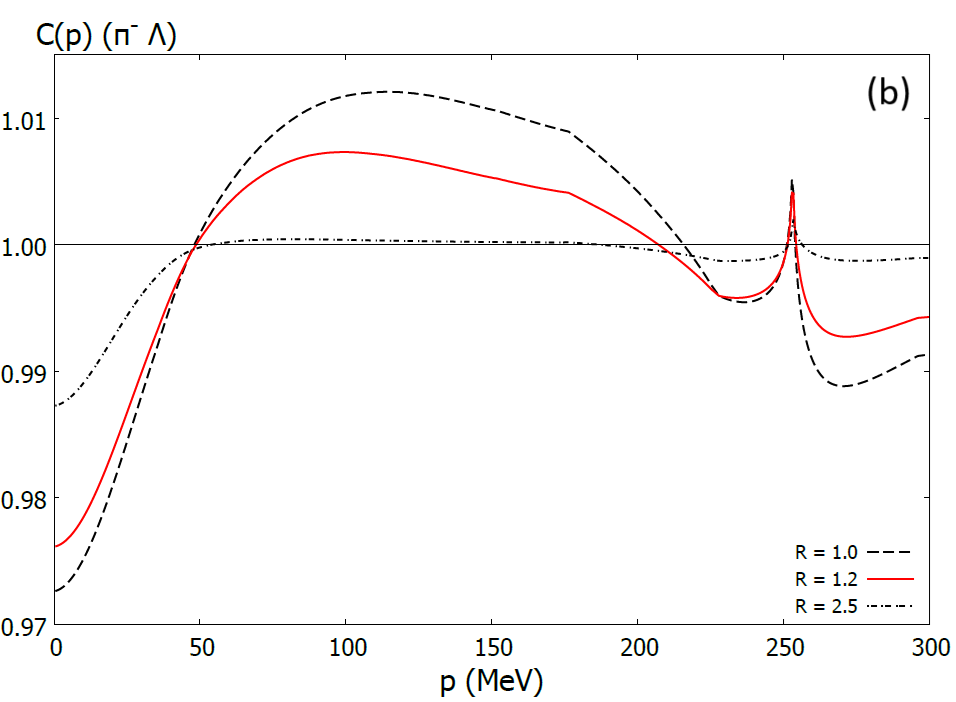}
    \end{minipage}
    \caption{a) Sequential contributions of the coupled channels to the $\pi^-\Lambda$ CF. b) $\pi^-\Lambda$ CF for different values of the source radii, $R=1.0$, $1.2$ and $2.5$ fm.}
    \label{fig:pi-lambda2}
\end{figure}

In Fig. \ref{fig:pi-lambda2}a the cc contributions to the $\pi^-\Lambda$ CF are depicted. In contrast to the $K^-p$ case, here the elastic channel is sufficient to produce the full CF at low momenta. As the momentum increases, the inelastic contributions become more relevant, specially around the $K^-n$ opening, where the full cusp structure is only achieved when including the $K^-n$ channel. Finally, in  Fig. \ref{fig:pi-lambda2}b the $\pi^-\Lambda$ CF is shown for three different source sizes: $R=1.0$, $1.2$ and $2.5$ fm. The CF highly depends on the source, its strength being enhanced by smaller sizes. Therefore, one must account for the source uncertainties once the data is available in order to properly compare the theoretical predictions with experiment.

\section{Summary and conclusions}

In this work we study the femtoscopic CFs of MB pairs in the $S=-1$ sector employing a cc unitarized chiral model up to NLO. When considering the experimental source size uncertainties, the calculated $K^-p$ CF is in good agreement with the ALICE experimental data. This is in contrast with the claim made by the same collaboration that the chirally motivated models should be revisited in order to reproduce the femtoscopic $K^-p$ data. We also give a novel prediction of the $\pi^-\Lambda$ CF. This channel will allow us to access the $K^-p$ subthreshold energy range of the $S=-1$ MB interaction and test the theoretical model once the corresponding experimental CF is available.


\begin{thebibliography}{99}












\bibitem{Alice2020} S. Acharya et al. (ALICE), Eur. Phys. J. C 83, 340 (2023)

\bibitem{Oset:onshell} E. Oset and A. Ramos, Nucl. Phys. A 635, 99 (1998)

\bibitem{Hyodo:onshell} T. Hyodo and D. Jido, Prog. Part. Nucl. Phys. 67, 55 (2012)

\bibitem{Oller:dimensional} J. A. Oller and U. G. Meissner, Phys. Lett. B 500, 263 (2001)

\bibitem{Feijoo:BCN} A. Feijoo, V. Magas, and A. Ramos, Phys. Rev. C 99, 035211 (2019)

\bibitem{Torres:coulomb} J. Torres-Rincón, A. Ramos, and L. Tolos, Phys. Rev. D 108, 096008 (2023).

\bibitem{Fabbietti} Fabbietti, L., Sarti, V. M., and Doce, O. V., Annu. Rev. Nucl. Part. Sci. 71:377-402 (2021)

\bibitem{Vidaña:fdo} I. Vidaña, A. Feijoo, M. Albaladejo, J. Nieves, and E. Oset, Physics Letters B 996, 138201 (2023).

\end{thebibliography}
\end{document}